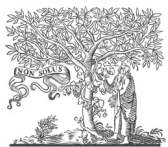
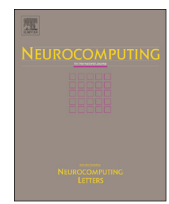

# Magnetic Resonance Fingerprinting with compressed sensing and distance metric learning

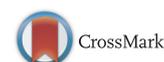

Zhe Wang [a], Hongsheng Li [a], Qinwei Zhang [b], Jing Yuan [b], Xiaogang Wang [a,*]

[a] Department of Electronic Engineering, The Chinese University of Hong Kong, Shatin, Hong Kong
[b] Department of Imaging and Interventional Radiology, the Chinese University of Hong Kong, Shatin, Hong Kong

ABSTRACT

Magnetic Resonance Fingerprinting (MRF) is a novel technique that simultaneously estimates multiple tissue-related parameters, such as the longitudinal relaxation time $T_1$, the transverse relaxation time $T_2$, off resonance frequency $B_0$ and proton density, from a scanned object in just tens of seconds. However, the MRF method suffers from aliasing artifacts because it significantly undersamples the k-space data. In this work, we propose a compressed sensing (CS) framework for simultaneously estimating multiple tissue-related parameters based on the MRF method. It is more robust to low sampling ratio and is therefore more efficient in estimating MR parameters for all voxels of an object. Furthermore, the MRF method requires identifying the nearest atoms of the query fingerprints from the MR-signal-evolution dictionary with the $\mathcal{L}_2$ distance. However, we observed that the $\mathcal{L}_2$ distance is not always a proper metric to measure the similarities between MR Fingerprints. Adaptively learning a distance metric from the undersampled training data can significantly improve the matching accuracy of the query fingerprints. Numerical results on extensive simulated cases show that our method substantially outperforms state-of-the-art methods in terms of accuracy of parameter estimation.

© 2015 Elsevier B.V. All rights reserved.

## 1. Introduction

Quantitative multiparametric acquisition in magnetic resonance imaging has long been the goal of research because it provides means of evaluating pathology using absolute tissue characteristics rather than contrast-based approaches [1]. It involves quantification of longitudinal relaxation time $T_1$, transverse relaxation time $T_2$, off resonance frequency $B_0$, proton density and other relevant parameters at each voxel of the scanned object. In most previous work, $T_1$ and $T_2$ are determined in separate scans [2–7]. Some recent methods can simultaneously estimate several parameters [8–10] but are restricted to only a limited set of parameters.

The Magnetic Resonance Fingerprinting (MRF) method recently proposed by [11] has the potential to quantitatively examine more than 4 magnetic resonance parameters simultaneously. The MRF method is based on the Inversion recovery-balanced SSFP (IR-bSSFP) [12] sequences. It has been reported that MRF outperforms the widely used DESPOT1 and DESPOT2 [7] methods for $T_1$ and $T_2$ estimation. It can also be used to directly estimate the combination proportions of different types of tissues at every single voxel. This may lead to new diagnostic methodologies.

The key idea of the MRF method is similar to matching a person's fingerprint to a database: once a match is made, additional information about the person can be obtained simultaneously. The MRF method generates unique signal evolutions by scanning a slice of the object for $T$ times with randomized system-related parameters. After applying the inverse Fourier transformation, the $T$-dimensional vector at every voxel location represents its characteristic signal evolution and is called its Magnetic Resonance Fingerprint [11]. Different tissues (such as white matter, gray matter, and cerebrospinal fluid) are assumed to have their own unique magnetic resonance fingerprints. These fingerprints can be easily distinguished by matching them to a predefined dictionary, which is generated using the well-known Bloch equation. The dictionary can be seen as a natural discretization of the Bloch response. It contains fingerprints of all foreseeable combinations of materials and system-related parameters. Each fingerprint corresponds to a vector of parameters to be estimated (such as $T_1$, $T_2$, $B_0$ and proton density). A nearest-neighbor based method is used to select the dictionary atom that best represents the observed fingerprint of a query voxel. All the magnetic resonance parameters corresponding to this dictionary atom can then be retrieved simultaneously. In this way, a set of MR parameters are estimated at every voxel location. The same procedure can be repeated to obtain MR parameter maps of all slices of the scanned object.

* Corresponding author.
*E-mail addresses:* lihongsheng@gmail.com (H. Li),
xgwang@ee.cuhk.edu.hk (X. Wang).

However, the MRF method still suffers from two problems: (1) in order to balance the accuracy and the scanning time, the MRF method significantly undersamples data in the k-space. Thus the reconstructed images exhibit extreme aliasing artifacts, which propagate to the estimated MR parameter maps. (2) The MRF method selects the best atom whose parameters are closest to the query fingerprint with the $\mathcal{L}_2$ distance. However, we observed that the $\mathcal{L}_2$ distance is not always appropriate for retrieving the correct fingerprint from the dictionary.

Recent developments in compressed sensing (CS) theory [13,14] show that it is possible to reconstruct signals from highly undersampled data, which provides plausible solutions for the first problem. So far, CS has been successfully applied to various domains in medical imaging, e.g., MR Imaging [15,16], shape modeling [17–19] and optical coherence tomography denoising [20].

Methods in [15,21] performed optimization with $L_1$ and *TV* norm regularizations by the Conjugate Gradient decent algorithm. These two methods could effectively reconstruct MR images with a sampling ratio around 20 %. Other methods like $L_p$ quasi-norm ($p<1$) regularized optimization [22,23] tolerate lower compression ratios, but these non-convex algorithms do not always recover global optima and are relatively slower. Ref. [24] adaptively learned the sparsifying transform (dictionary) and thereby favoring higher sparsity and consequently higher sampling ratios. Their reconstructions can achieve higher undersampling ratios with tolerable errors. However, all these algorithms lead to aliasing artifacts if they are applied to the MRF method with a sampling ratio of only around 3%.

To our knowledge, two previous methods [25,26] were proposed to integrate a CS algorithm into the MRF framework. Ref. [25] proposed to apply CS to reconstruct the image at each sampling time. But the sampling ratio cannot be less than 70%. Ref. [26] adopted a CS solution based on the iterative projection algorithm by [27] which imposes consistency with the Bloch response manifold. At each iteration, every voxel would be replaced by its nearest atom in the dictionary. Then the whole image at each sampling time was updated by the Projected Landweber Algorithm (PLA). This method is called BLIP (BLoch response recovery via Iterated Projection), and is efficient and effective in removing the aliasing artifacts.

Moreover, all the previous works [11,25,26] used inner-product for calculating the similarities between the query fingerprint and the dictionary atoms, which is equivalent to using the $\mathcal{L}_2$ distance as the distance metric. However, we observe that if the distance metric is learned in a supervised manner, then the performance of the nearest-neighbor based dictionary matching can be significantly improved.

In this work, we propose a compressed sensing framework for simultaneously estimating multiple MR parameter maps with distance metric learning. Instead of treating each voxel individually, we assume that each image is sparse in some transform domain. The problem of estimating MR parameter maps is then formulated as a compressed sensing problem, where we make use of the spatial information of the image sequence. For each voxel, its fingerprint is then matched to its nearest atom in a predefined dictionary with a learned distance metric. Such a learned metric is more accurate in MR fingerprint matching. Furthermore, a novel sampling strategy based on Cartesian sampling is proposed. Our strategy makes the aliasing noise as incoherent as possible with the fingerprint itself, thus making it easier to be removed. Extensive experiments were conducted on simulated MR images to evaluate the performance of the proposed method. Numerical results show that it outperforms state-of-the-art methods in estimating multi-parametric MR maps of scanned objects.

In real scenarios, the ground truth MR parameter maps for distance metric learning can be obtained by applying standard MR imaging approaches to volunteers or phantoms. The learned metric can then be used for future scans under the same experimental setting.

Our main contribution is three-fold: (1) we propose a compressed sensing framework based on MRF that is more robust to estimate multiple MR parameter maps of a scanned object at low sampling ratios. It makes use of the spatial information of the image sequence and is therefore accurate in estimating the MR parameters when the sampling ratio is very low. (2) We improve the accuracy of the dictionary matching process by replacing the $\mathcal{L}_2$ distance with a learned distance metric. The proposed metric can be learned in a supervised manner. (3) In order to make the undersampling errors and the MR fingerprints as incoherent as possible, we design a novel sampling strategy with which the sampling mask at time $t$ is conditional on the one at time $t-1$. It generates aliasing noise that is easier to be removed by dictionary matching.

## 2. Methodology

In this work, we propose a novel framework to simultaneously estimate multiple MR parameters for every voxel of a scanned object based on the MRF method. In Section 2.1, the MRF method and notation is introduced. In Section 2.2, we introduce a compressed sensing framework for MRF to reduce errors. In Section 2.3, we present adaptively learning a distance metric for dictionary matching. A novel sampling strategy is proposed in Section 2.4 which further removes the aliasing noise.

### 2.1. Magnetic Resonance Fingerprinting (MRF)

The key underlying assumption in MRF is that different materials or tissues have their own unique signal evolutions or fingerprints. The magnetization at a given voxel location at time $t$ depends on its magnetic resonance parameters and the system-related parameters, including the flip angle *FA*, repetition time *TR* and others, at time $t-1$. For illustration purposes, we explain the estimation of MR parameter maps of only a single slice in Section 2.

Let $X \in \mathbb{C}^{N \times T}$ denote multiple scans of one slice of the object of interest, where $N$ is the total number of voxels in the slice and $T$ is the sequence length. Let $X_t^i \in \mathbb{C}$ denote the $i$th voxel of the scanned slice at time $t$, $X^i \in \mathbb{C}^{1 \times T}$ denote the signal evolution or fingerprint at voxel $i$ at all times, and $X_t \in \mathbb{C}^{N \times 1}$ denote the scanned image of the slice at time $t$.

Given the initial magnetization, the signal evolution or fingerprint at voxel $i$ can be written as

$$X^i = \rho_i \mathcal{B}(\theta_i; FA, TR), \tag{1}$$

where $\rho_i$ is the proton density – one of the magnetic resonance parameters to be estimated, $\theta_i$ is the collection of other magnetic resonance parameters at voxel $i$, and $\mathcal{B}$ is the Bloch equation dynamics.

Since the possible range of $\theta_i$ of the object is known in advance, we densely sample each MR parameter and use the Bloch equation to create the dictionary $D \in \mathbb{C}^{K \times T}$, where $K$ is the number of dictionary atoms. Each dictionary atom is normalized so that $\|D^k\|_2 = 1$, for $k = 1, 2, \ldots, K$. The same set of system-related parameters *FA* and *TR* is used for both creating the dictionary and obtaining the scanning data $X$. Given a query fingerprint, it is matched to its nearest atom in the predefined dictionary with the $\mathcal{L}_2$ distance. The index of the nearest dictionary atom for the fingerprint $X^i$ is denoted as $\tilde{k}_i$, and is obtained as

$$\tilde{k}_i = \underset{k}{\arg\min} \, \|X^i / \|X^i\|_2^2 - D^k\|_2 \tag{2}$$

$$\tilde{k}_i = \underset{k}{\arg\max} \, \left\{ real \langle X^i / \|X^i\|_2^2, D^k \rangle \right\}, \tag{3}$$

where *real* is the operation to extract the real part of a complex number and $\langle \cdot, \cdot \rangle$ is the inner product operation. The corresponding parameters of the fingerprint $X_i$ are obtained as

$$\tilde{\theta}_i = \Gamma(\tilde{k}_i), \tag{4}$$

where $\Gamma$ retrieves the MR parameters based on the dictionary index. The proton density at voxel $i$ is then estimated as

$$\tilde{\rho}_i = \max\left\{ real\langle X^i/\|X^i\|_2^2, D^{\tilde{k}_i}\rangle, 0 \right\}, \tag{5}$$

where the max operation is applied to remove unacceptable negative values.

Only a limited portion of the k-space data is collected at each time frame. Due to the undersampling, the image directly obtained by the inverse Fourier transform will be contaminated with strong aliasing artifacts.

### 2.2. Compressed sensing for Magnetic Resonance Fingerprinting (CSMRF)

In this section, we propose a compressed sensing framework based on MRF for simultaneously estimating multiple tissue-related parameters with tolerance to very low sampling ratios.

For some fingerprint with minor aliasing noise, once it is replaced by its nearest atom in the dictionary by Eq. (3), the undersampling error at that voxel location is already eliminated. This has been shown in [11].

However, the exact matching may fail because the query fingerprint is impaired by significant undersampling errors. Instead of treating the aliasing artifacts as random noise as in [11], we treat it as the leakage of energy caused by undersampling, which can be estimated by the theory of compressed sensing. Under this assumption and some additional conditions, the missing signals can be perfectly recovered. The problem of reconstructing undersampled k-space data can be formulated as a compressed sensing problem:

$$\min_{X_t} \quad \|\Phi X_t\|_1$$
$$\text{s.t.} \quad \|\mathcal{F}_u(X_t) - Y_t\|_2^2 < \epsilon, \tag{6}$$

where $F_u$ is the Fourier transform operator with our proposed sampling mask (the details will be described in Section 2.4), and $Y_t$ is the k-space measurement at time $t$. Minimizing the $\|\cdot\|_1$ term forces the image $x$ to be sparse in some transform domain $\Phi$. In this work, we assume that the image is sparse in 2 domains, i.e., (1) the wavelet domain with Daubechies filters of 4 scales and (2) the finite difference domain. The wavelet transform term often results in the removal of high frequency noise-like patterns, while the finite difference term favors solutions that are piecewise smooth [15]. The $\|\cdot\|_2^2$ term requires the image $X_t$, when transformed back to the k-space, being consistent with the measurement $Y_t$. The $\epsilon$ controls the fidelity of the reconstruction to the measured data. The optimization problem (6) can be rewritten in a Lagrangian form:

$$\hat{X} = \operatorname*{argmin}_{X_t} \|\mathcal{F}_{\hat{u}}(X_t) - Y_t\|_2^2 + \alpha \|\Phi X_t\|_1, \tag{7}$$

where $\alpha$ is a weight parameter. This problem can be solved by the Conjugate Gradient algorithm [15].

Optimizing the compressed sensing problem can be considered as utilizing the spatial information to remove aliasing artifacts in the reconstructed images. After that the denoised fingerprint $\hat{X}^i$ is matched to the nearest dictionary atom with a learned Mahalanobis distance metric $A$, which can be written as

$$\hat{k}_i = \operatorname*{argmin}_{k} \|\hat{X}^i/\|\hat{X}^i\|_2^2 - D^k\|_A, \tag{8}$$

The learning of the metric $A$ is detailed in the next section. The learned distance metric captures important dimensions of the fingerprints in the temporal domain, and is more accurate than the $\mathcal{L}_2$ distance used in previous work [11,26]. In this way, the temporal information is also used in our framework.

Given the dictionary index $\hat{k}_i$, the tissue-related parameters $\hat{\theta}$ can be retrieved as

$$\hat{\theta} = \Gamma(\hat{k}_i), \tag{9}$$

and the proton density is thus

$$\hat{\rho}_i = \max\left\{ real\langle \hat{X}^i, D^{\hat{k}_i}\rangle_2^2, 0 \right\}. \tag{10}$$

To sequentially solve the two objective equations (7) and (8), we propose an algorithm summarized in Algorithm 1. In the initialization step (line 2–4), all the images $X$ are obtained by applying the inverse Fourier transform on the measurements $Y$ in the k-space. In the spatial update step (line 6–8), each image is reconstructed by optimizing Eq. (7). In the temporal update step (line 10–14), we match each fingerprint to its nearest atom in the dictionary with the learned metric.

**Algorithm 1.** Compressed sensing for Magnetic Resonance Fingerprinting with metric learning (CSMRF+ML)

---
**Input**: Measurements $Y$, Dictionary $D$, Learned metric $A$
**Output**: Tissue-related parameters $\hat{\rho}$, $\hat{\theta} = \{\hat{T}_1, \hat{T}_2, \hat{B}_0\}_i$
1  // Initialization:
2  **for** $t = 1$ to $T$ **do**
3  $\quad X_t = \mathcal{F}_{\hat{u}}^H(Y_t)$
4  **end**
5  // Update each image
6  **for** $t = 1$ to $T$ **do**
7  $\quad$ Update $X_t$ by optimizing Eq. (7).
8  **end**
9  // Update each fingerprint
10 **for** $i = 1$ to $N$ **do**
11 $\quad \hat{k}_i = \operatorname{argmin}_k \|X^i/\|X^i\|_2^2 - D^k\|_A$
12 $\quad \hat{\theta}_i = \Gamma(\hat{k}_i)$
13 $\quad \hat{\rho}_i = max\left\{real\langle X^i, D^{\hat{k}_i}\rangle, 0\right\}$
14 $\quad X^i = \hat{\rho}_i D^{\hat{k}_i}$
15 **end**
---

### 2.3. Distance metric learning for MR Fingerprint matching

To match a query fingerprint to a dictionary atom as shown in Eq. (8), we propose to learn a Mahalanobis distance, with which the query fingerprint is closest to its corresponding atom in the pre-defined dictionary, to replace the $\mathcal{L}_2$ distance. This is because the $\mathcal{L}_2$ distance might not be a proper metric for measuring the dissimilarity between MR Fingerprints. In Fig. 1(a), we show such an example where the $\mathcal{L}_2$ distance fails to retrieve correct MR parameters for a query fingerprint corrupted by strong aliasing noise. For the query fingerprint (red), its nearest dictionary atom (green) with the $\mathcal{L}_2$ distance is quite different from its ground truth atom (blue).

The Mahalanobis distance between normalized fingerprints $\hat{X}^i$ (i.e., $\|\hat{X}^i\|_2 = 1$) and $D^j$ is defined as

$$\mathbf{d}_A^2(\hat{X}^i, D^j) = \|\hat{X}^i - D^j\|_A^2 \tag{11}$$

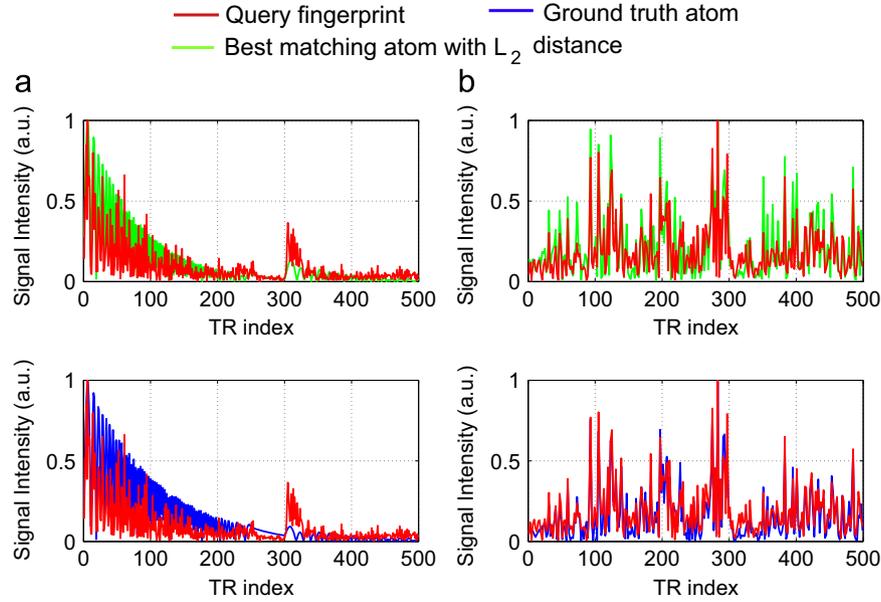

**Fig. 1.** Simulated signal evolutions corrupted by strong aliasing noise. The experimental settings can be found in Section 3. (a) For the query fingerprint (red), its nearest dictionary atom $D_{min}$ (green) with the $\mathcal{L}_2$ distance is quite different from its ground truth atom $D_{gt}$ (blue). (b) After a distance metric is learned, the fingerprint and dictionary atoms are transformed such that the ground truth atom $D_{gt}$ (blue) is now nearest to the query fingerprint, while the nearest dictionary atom $D_{min}$ in (a) (green) is now quite different from the query fingerprint. The plots are normalized w.r.t. their maximum values. (Best viewed in color). (For interpretation of the references to color in this figure caption, the reader is referred to the web version of this paper.)

$$= \left(\hat{X}^i - D^j\right)^T A \left(\hat{X}^i - D^j\right) \qquad (12)$$

$$= \left(\hat{X}^i - D^j\right)^T W^T W \left(\hat{X}^i - D^j\right)$$

$$= \|W\hat{X}^i - WD^j\|_2^2, \qquad (13)$$

where the superscript $T$ denotes the transpose, $A$ is the learned positive semi-definite matrix and $W$ is the transformation matrix. If $A$ is the unit matrix, then the distance degenerates to the $\mathcal{L}_2$ distance. Note that since most distance metric learning algorithms deal with real signals, we concatenate the real parts and the imaginary parts of all training fingerprints. For illustration purposes, we still use the notation $\hat{X}^i$ and $D^j$ in the remaining parts of the paper. Note that all the data is preprocessed if the Mahalanobis distances between them is calculated.

In this work, we adopt the Relevant Component Analysis (RCA) [28] algorithm and found that it is superior than other distance metric learning algorithms in our framework. The RCA requires training samples (fingerprints) and their labels to learn an optimal distance metric between them. The training fingerprints consist of fingerprints from the image sequence and their corresponding dictionary atoms. Labels are assigned to the training fingerprints in the following way. Fingerprints corresponding to the same dictionary atom are given the same label. The same label is also assigned to the corresponding dictionary atom. Notice that although neighboring dictionary atoms may also be good candidates for the query fingerprint, we do not merge their labels. Dictionary atoms that have no corresponding fingerprints are not included in the training samples. Those atoms may represent materials or tissues rarely exist in the scanned object.

Let $M$ denote the number of training fingerprints with $L$ different labels, $\{P_{ji}\}_{i=1}^{n_j}$ denote the fingerprints in the chunklet $j$, where $n_j$ is the number of fingerprints in the $j$th chunklet. A chunklet means a subset of fingerprints that are known to share the same label.

The objective function of the RCA is formulated as

$$\max_A \log|A| \quad \text{s.t.} \quad \frac{1}{M}\sum_{j=1}^{L}\sum_{i=1}^{n_j}\|P_{ji} - \overline{P}_j\|_A^2 \leq 1, \qquad (14)$$

where $\overline{P}_j$ is the mean of the $j$th chunklet.

Multiplying a solution $A$ by a constant larger than 1 increases the objective value as well as the constrained sum. Therefore, the solution is obtained at the boundary of the feasible region, where the inequality constraint becomes an equality, i.e.,

$$\max_A \log|A| \quad \text{s.t.} \quad \frac{1}{M}\sum_{j=1}^{L}\sum_{i=1}^{n_j}\|P_{ji} - \overline{P}_j\|_A^2 = 1. \qquad (15)$$

Solving the equality constraint leads to the solution, which is linear in $A$. Let the within chunklet covariance matrix $C$ be

$$C = \frac{1}{M}\sum_{j=1}^{L}\sum_{i=1}^{n_j}(P_{ji} - \overline{P}_j)(P_{ji} - \overline{P}_j)^T. \qquad (16)$$

The optimal transformation matrix is thus calculated as $W = C^{-1/2}$, which has large weights on relevant dimensions and small weights on irrelevant dimensions.

As shown in Fig. 1(b), after a distance metric is learned for the example in Fig. 1(a), the fingerprint and dictionary atoms are transformed such that the query fingerprint is now most similar to the ground truth atom.

Once the experimental parameters FA, TR and the undersampling pattern are determined, the learned distance metric can be used again for future scans together with them. Another observation is that the matrix have large entries in the main diagonal, which agrees with the previous works [11,26] that the $\mathcal{L}_2$ distance is also a benign distance metric choice.

In practice, the distance metric can be learned in advance. First, the MR parameter maps of either phantoms or volunteers can be obtained by standard MR imaging methods. Then with fixed imaging-related settings, such as FA, TR, and sampling strategy, the MR fingerprints can be collected and used for training the distance metric. The learned metric is applicable for later use under the same experimental settings.

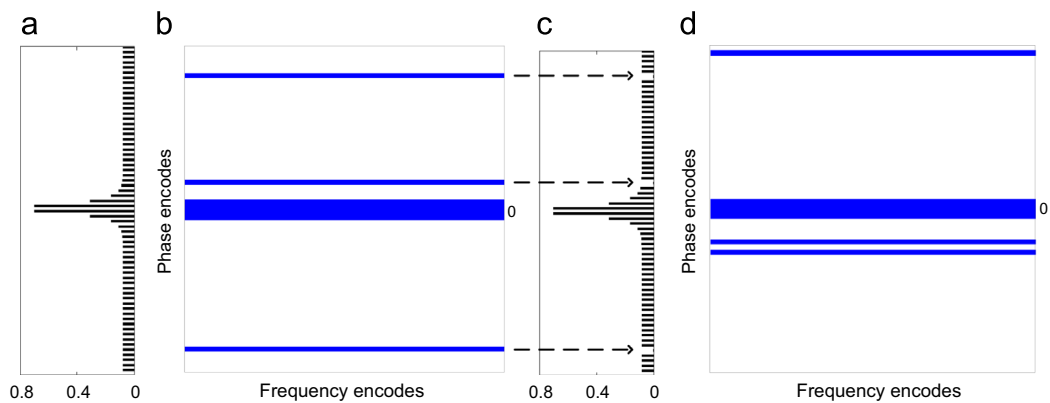

**Fig. 2.** Illustration of how the sampling masks are generated at two consecutive times frames by the proposed sampling strategy. (a) The $(t-1)$th sampling probability. (b) The $(t-1)$th sampling mask generated using the sampling probability in (a). (c) The $t$th sampling probability. The dashed arrows point to the probabilities that are forced to be 0 because they are already sampled in (b). (d) The $t$th sampling mask generated using the sampling probability in (c).

### 2.4. The time-dependent sampling strategy

We propose an undersampling strategy based on Cartesian sampling, in which only the phase encodes are randomly sampled. Each phase encode line is represented by a row in the k-space matrix. Due to the sampling mechanism in MRI, unlike frequency encoding, each phase encode line has to be entirely sampled, thus we cannot achieve the random sampling in two directions which is assumed to be ideal for compressed sensing. The method [15] on MR image reconstruction has shown that the k-space should be sampled more at the lower frequencies because most energy is concentrated around the k-space origin. At the same time, the reconstruction of MR fingerprints requires the aliasing noise at each time to be as incoherent as possible.

Here we propose a sampling strategy that takes both requirements into consideration. A sampling probability $sp_1$ at time 1 is first initialized following [15] such that it samples more near the k-space origin. More specifically, the probability of sampling a row scales according to a power of distance from the k-space (see Fig. 2 (a) for illustration). The sampling of each row of the mask at time $t-1$ then follows a binomial distribution parametrized by the probability value at that location. The sampling probability $sp_t$ at time $t$ is conditional on the mask at time $t-1$. If some row on the $(t-1)$th mask has been sampled, then the entry of the $t$th probability corresponding to the same location is set to zero, except for the $c$ rows nearest to the k-space center. Thus we force the consecutive sampling masks to be as different as possible, while they still sample more data from the lower frequencies. The sampling probability for the $i$th row in the k-space at time $t$ is defined as

$$sp_t(i) = \begin{cases} sp_1(i), & i \in C \cup \overline{M_{t-1}} \\ 0 & \text{otherwise} \end{cases} \quad (17)$$

where $sp_t$ is the probability at time $t$, $C$ is the index set of the $c$ rows to be kept near the k-space origin, $M_{t-1}$ is the index set of the rows that are sampled in the $(t-1)$th mask, and $\overline{M_{t-1}}$ is the complementary set of $M_{t-1}$. If $c$ equals the total number of rows, then all the sampling masks are independent. By controlling the $c$ value we can balance between the sampling of low frequency parts and high frequency parts of the k-space data.

## 3. Results

### 3.1. Experimental settings

#### 3.1.1. Data
We tested our proposed method on anatomical brain phantoms [29] that were downloaded from the BrainWeb repository [30]. The sizes of the original phantoms are $434 \times 362$ with 362 slices. The 174th slice was used for training the distance metric. Slices 175–184 were used for testing the accuracy of our method and the compared methods. Both training and testing brain phantoms were rescaled and padded to $256 \times 256$ (Fig. 3).

The phantoms were restricted to contain 7 material components as listed in Table 1. In order to simulate more realistic images, we added variations to the parameter maps. The $T_1$ map was added by noise drawn from a standard uniform distribution on an interval of 0–50 ms. The $T_2$ map was added with noise drawn from a standard uniform distribution on an interval of 0–10 ms. The $B_0$ map was added with noise drawn from a standard uniform distribution on an interval of 0–10 Hz.

#### 3.1.2. Dictionary generation
The four parameter maps were retrieved simultaneously from each $256 \times 256$ slice using our method. The parameter maps were restricted so that $T_1$ ranged from 300 to 4700 ms, $T_2$ ranged from 45 to 600 ms and $B_0$ ranged from $-100$ to 200 Hz. The size of the designed dictionary was $48 \times 53 \times 41 \times T$, where $T$ is the sequence length.[1] The entries for $T_1$ were sampled from 300 to 1000 ms with an increment of 30 ms, from 1000 to 2500 ms with an increment of 100 ms, and from 1500 to 4700 ms with an increment of 300 ms. The $T_2$ values are sampled from 45 to 100 ms, 320 to 370 ms with an increment of 10 ms, from 110 to 320 ms, and from 380 to 630 ms with an increment of 50 ms. The $B_0$ values are sampled from $-200$ to 200 Hz with an increment of 10 Hz. To make the simulation more close to the real scenarios, both the parameter maps and the dictionary were designed so that no exact fingerprint match can be found.

#### 3.1.3. Sequence setting
We simulated the signal evolutions with the IR-bSSFP sequence using a randomized series of flip angles and repetition times of 10 ms. While we also experimented with randomized repetition times, no significant performance change was observed. Let $\eta$ be the noise term sampled from a Gaussian distribution with a standard deviation of 5 ms. The flip angles, $FA$, are calculated as a series of repeating sinusoidal curves added with Gaussian random

---

[1] The code for generating the dictionary and fingerprint matching with $\mathcal{L}_2$ distance is obtained from supplementary information of [11]: http://www.nature.com/nature/journal/v495/n7440/extref/nature11971-s1.pdf

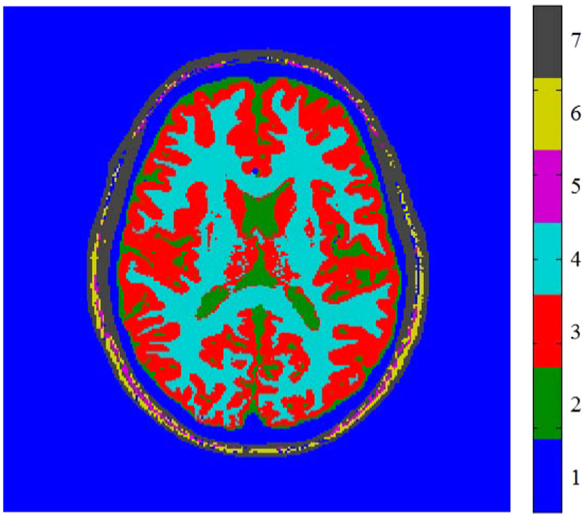

**Fig. 3.** The segmented anatomical brain phantom [29] colored by index: 1=Background, 2=CSF, 3=Grey Matter, 4=White Matter, 5=Fat, 6=Muscles, 7=Muscles/Skin. (For interpretation of the references to color in this figure caption, the reader is referred to the web version of this paper.)

**Table 1**
Tissue types used from segmented brain phantom.

| Tissue | Index | $T_1$ | $T_2$ | $B_0$ | Density |
|---|---|---|---|---|---|
| Background | 1 | 0 | 0 | 0 | 0 |
| CSF | 2 | 4231 | 572 | 185 | 1 |
| Gray matter | 3 | 833 | 86 | −30 | 0.86 |
| White matter | 4 | 500 | 55 | −70 | 0.77 |
| Fat | 5 | 350 | 70 | −80 | 0.7 |
| Muscle | 6 | 900 | 47 | −40 | 1 |
| Muscle/skin | 7 | 2269 | 329 | 75 | 1 |

noise, i.e.,

$$FA(t) = \begin{cases} 10 + \sin(2\pi t/500) \times 50 + \eta, & 0 < t \leq 250 \\ 10, & 250 < t \leq 300 \\ 5 + \sin(2\pi/200 \times 25) + \eta, & 300 < t \leq 500 \end{cases}$$

#### 3.1.4. Evaluation metrics

The quality of the MR parameter estimation is quantified using the Peak-Signal-to-Noise-Ratio (PSNR) in decibels and the Structural SIMilarity (SSIM) index [31]. When computing PSNR, the MR parameter maps are first normalized to the range of [0,255]. PSNR is then computed as the ratio of the peak intensity value of the ground truth to the Mean Square Error (MSE) reconstruction error relative to the ground truth. The SSIM index is developed as a complementary approach to the traditional metrics based on error-sensitivity. It has been shown to be more consistent with human eye perception [31]. Unlike PSNR which estimates perceived errors, SSIM considers image degradation as perceived change in structural information. The SSIM index is a decimal value between −1 and 1, and value 1 can only be reached when two images are identical.

#### 3.1.5. Compared methods

Our proposed CSMRF+ML was able to recover the MR parameter maps accurately. We compared it with the MRF [11] and BLIP [26] methods. For simplicity and better comparison, the proposed sampling strategy is used for MRF. For BLIP, the uniform sampling strategy based on EPI is used as in [26], i.e., the rows in k-space are undersampled by a factor $p$ with random shifts across time. In some experiments, we also reported the performance of the oracle estimator, which was obtained by matching the fully sampled image sequence data to the nearest dictionary atoms using the $\mathcal{L}_2$ distance metric. Because the oracle estimator samples all the data, it should always achieve the best estimation results if the dictionary is correctly created. In this way, we could differentiate the errors caused by the Bloch response discretization and those by the other factors.

### 3.2. Overall performance and comparisons with existing methods

The overall performance of our method and the compared MRF [11] and BLIP [26] methods is reported in this section. We set the sequence length $T$ to 500, with a sampling ratio of 6.25% (16 rows out of 256). The $c$ value in Eq. (19) is chosen to be 6. All the experiments were performed on 10 different slices, and the averaged results were reported in Table 2 (also see the visual comparisons in Figs. 4–7). As expected, the oracle estimator achieved the best result among all methods because it does not undersample the k-space data. Our proposed method CSMRF+ML performs best compared to MRF and BLIP on the estimation of $T_1$, $T_2$ and off resonance frequency maps, while the accuracy improvement of proton density maps is slightly lower than BLIP. This is because the proton density depends on the product of the query fingerprint and the matched dictionary atom, although our methods can find better matches than MRF, the final result will still be affected by the shrinkage effect of undersampling, even after the density compensation is applied to the k-space. And since BLIP iteratively updates the fingerprints by alternatively projecting them to the Bloch response manifold and minimizing the reconstruction error, it can estimate the proton density map better. BLIP shows better accuracy than MRF. However, its performance is not stable and depends on the randomness of its sampling masks because the sampling strategy used by BLIP does not have constraints on sampling masks at consecutive times.

Figs. 4–7 show example estimated maps by different algorithms. MRF and CSMRF+ML share the same set of sampling masks by Eq. (19), while BLIP uses uniformly undersampling strategy based on EPI. We set BLIP to run 16 iterations, and no significant improvement is observed if more iterations of operations are performed. The result of MRF shows substantial aliasing artifacts. BLIP successfully removes most of the aliasing noise, while the estimated $T_1$, $T_2$ and off resonance frequency maps of CSMRF+ML exhibit almost no aliasing artifacts. The proton density map of CSMRF+ML is slightly overestimated due to the shrinkage effect.

### 3.3. Performance with noise

To evaluate the noise robustness of the proposed method, we added zero-mean complex Gaussian noise of standard deviation $\sigma = 0.5$ to the k-space of all the frames. An example of the fully sampled noisy frame at time 1 is shown in Fig. 8(a), and can be observed to be considerably noisy. The PSNR of the noisy image with respect to the reference is 19.1 dB. The distance metric is trained with another set of images contaminated by the same type of noise but with different random seeds. Here we show the estimated $T_2$ maps by MRF, BLIP and CSMRF+ML in Fig. 8(b)–(d). The results show that MRF is unable to effectively remove the noise. BLIP performs better than MRF and show a clearer reconstruction result. However, there still exists certain amount of noise. Our methods CSMRF+ML on the other hand generates a satisfactory result. Almost all the noise is eliminated except for some on the boundaries. The PSNR and SSIM of CSMRF+ML is 4.2 dB and 0.083 higher than BLIP, respectively, which shows that CSMRF+ML can perform better in the presence of reasonable amount of noise.

### 3.4. Evaluation on different components

In this section, the effect of each individual parameter is investigated. We compare CSMRF+ML and the state-of-the-art methods against different sampling ratios in Section 3.4.1, and against different sequence lengths $T$ in Section 3.4.2. In Section 3.4.3, we compare the proposed sampling strategy with a baseline strategy. The effect of different $c$ values of our strategy is also investigated. In Section 3.4.4, different distance metric learning algorithms are tested for CSMRF+ML.

**Table 2**
Quantitative results of the proposed algorithm and the state-of-the-art algorithms. The winning entries are marked in bold.

| Method | $T_1$ map | | $T_2$ map | | $B_0$ map | | Density map | |
|---|---|---|---|---|---|---|---|---|
| | PSNR | SSIM | PSNR | SSIM | PSNR | SSIM | PSNR | SSIM |
| Oracle estimator | 42.8 | 0.99 | 40.6 | 0.99 | 52.0 | 1.0 | 92.6 | 1.0 |
| MRF | 27.0 ± 0.54 | 0.95 ± 0.02 | 22.9 ± 0.32 | 0.86 ± 0.06 | 24.8 ± 0.51 | 0.89 ± 0.03 | 23.6 ± 0.23 | 0.87 ± 0.02 |
| BLIP | 30.2 ± 1.66 | 0.96 ± 0.12 | 26.8 ± 0.81 | 0.86 ± 0.07 | 28.2 ± 1.12 | 0.90 ± 0.05 | **28.6 ± 0.96** | 0.88 ± 0.04 |
| CSMRF+ML | **31.1 ± 0.87** | **0.99 ± 0.01** | **37.3 ± 0.76** | **0.99 ± 0.01** | **39.9 ± 0.64** | **0.99 ± 0.01** | 25.8 ± 0.46 | **0.92 ± 0.04** |

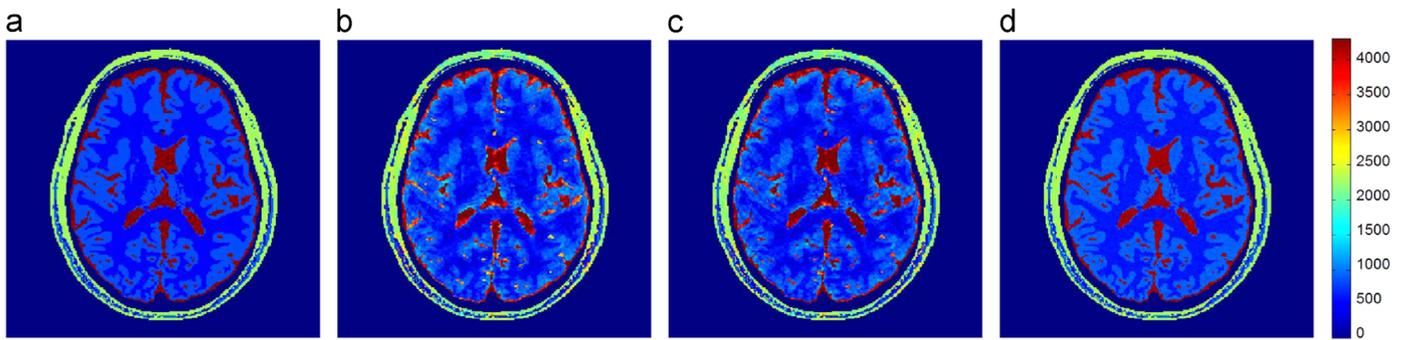

**Fig. 4.** Example estimated $T_1$ maps with a sampling ratio of 6.25%. (a) Ground truth. Results by (b) MRF [11], (c) BLIP [26] and (d) CSMRF+ML. All the images are displayed in the same color range. (For interpretation of the references to color in this figure caption, the reader is referred to the web version of this paper.)

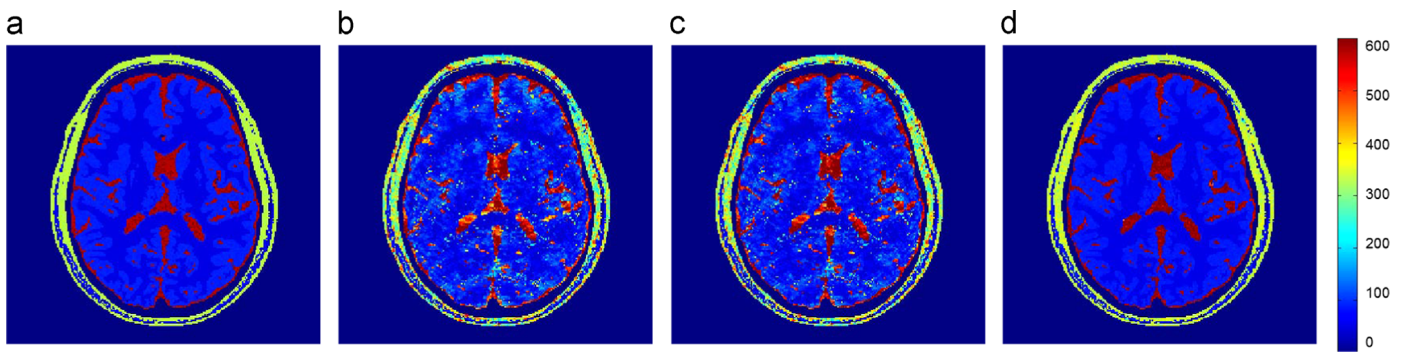

**Fig. 5.** Example estimated $T_2$ maps with a sampling ratio of 6.25%. (a) Ground truth. Results by (b) MRF [11], (c) BLIP [26] and (d) CSMRF+ML. All the images are displayed in the same color range. (For interpretation of the references to color in this figure caption, the reader is referred to the web version of this paper.)

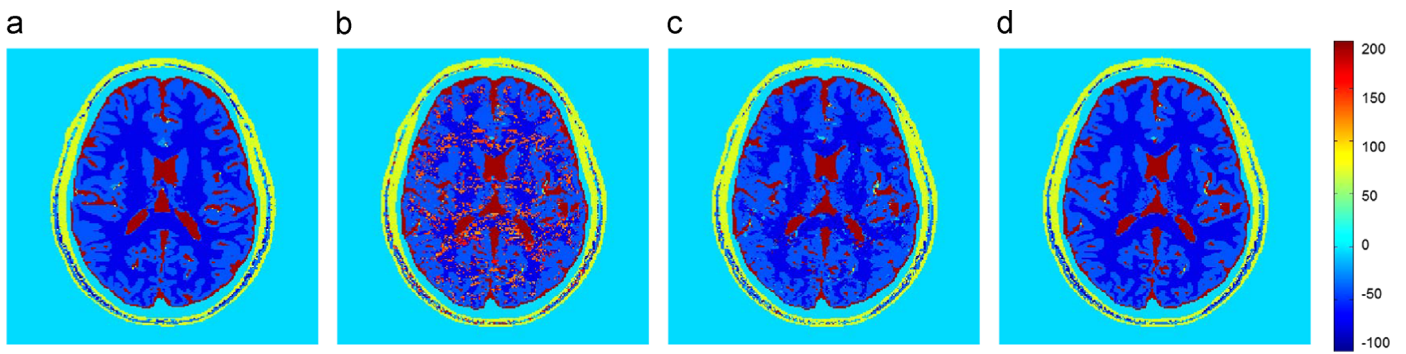

**Fig. 6.** Example estimated $B_0$ maps with a sampling ratio of 6.25%. (a) Ground truth. Results by (b) MRF [11], (c) BLIP [26] and (d) CSMRF+ML. All the images are displayed in the same color range. (For interpretation of the references to color in this figure caption, the reader is referred to the web version of this paper.)

### 3.4.1. Evaluation on sampling ratio

In this section, we evaluated the estimation accuracy of our algorithm with different sampling ratios. Recall that the images are of size $256 \times 256$. We experimented with sampling ratios of 3.13%, 3.91%, 4.69%, 5.49%, 6.25%, 7.03% and 7.81% (which are equivalent to 8, 10, 12, 14, 16, 18 and 20 rows per image), respectively. We report the PSNR of the estimated MR parameter maps vs. the sampling ratios here.

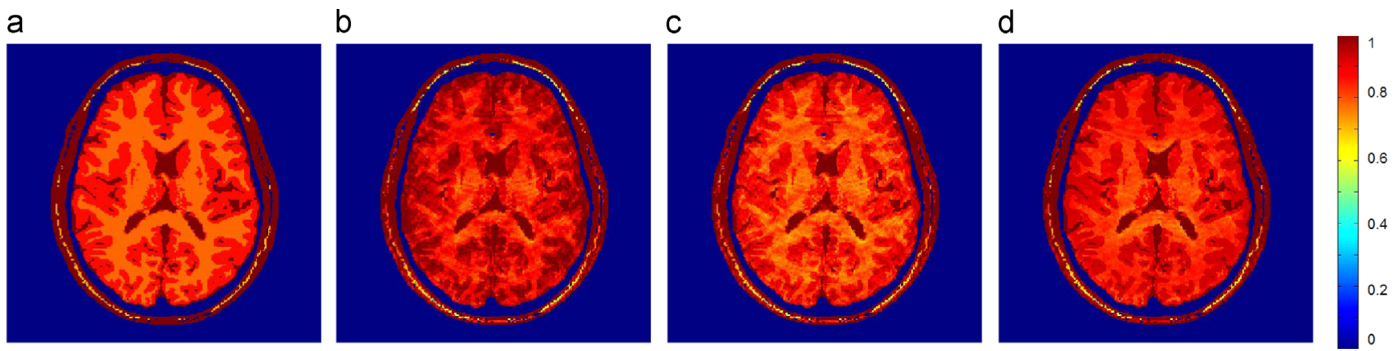

**Fig. 7.** Example estimated density maps with a sampling ratio of 6.25%. (a) Ground truth. Results by (b) MRF [11], (c) BLIP [26] and (d) CSMRF+ML. All the images are displayed in the same color range. (For interpretation of the references to color in this figure caption, the reader is referred to the web version of this paper.)

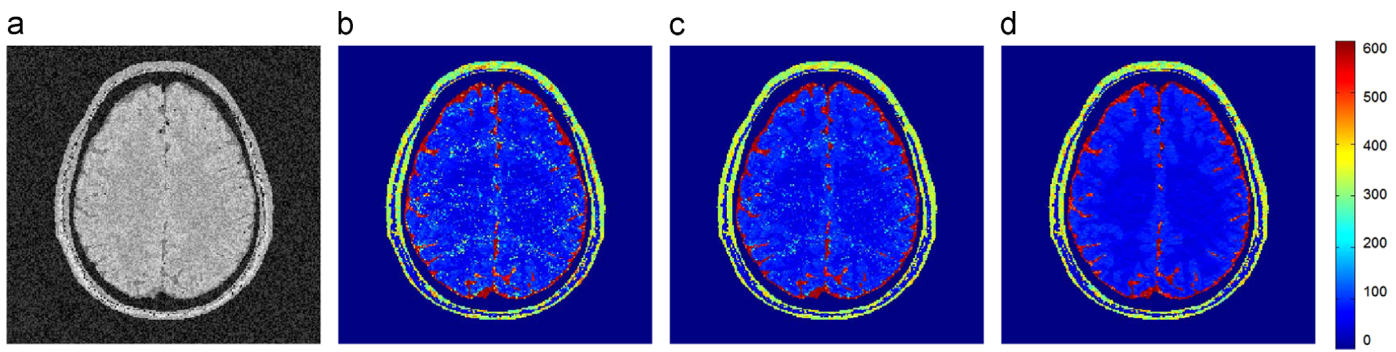

**Fig. 8.** (a) Fully sampled noisy frame at time 1. $T_2$ maps reconstructed by (b) MRF [11], (c) BLIP [26] and (d) CSMRF+ML. All the image are displayed in the same color range. (For interpretation of the references to color in this figure caption, the reader is referred to the web version of this paper.)

In Fig. 9, we show that the overall estimation accuracy increases as the sampling ratio increases for all the MR parameter maps. Interestingly, BLIP has a performance boost at sampling ratios of 3.13% and 6.25% (8 and 16 rows per image). This may be because the sampling strategy used by BLIP requires under-sampling the k-space uniformly with random shifts across time. Ideally, each row should have the same chance to be sampled during the whole process, which will maximize the randomness of different sampling masks. However, if the number of rows to be sampled is not divisible by the total number of rows (e.g. sampling 12 out of 256 rows), then some rows might never be sampled, which leads to degraded performance. Note that in [26], only sampling ratios of 6.25%, 12.5% and 25% are shown.

With metric learning and the proposed sampling strategy, CSMRF+ML always performs best. Since CSMRF+ML uses a non-uniform sampling strategy, it does not suffer from the problem described above. It can achieve satisfactory and stable reconstruction quality at arbitrary sampling ratios in our experiments.

### 3.4.2. Evaluation on sequence length

This experiment evaluated the performance of the proposed algorithm with different sequence lengths, which varied from 100 to 500. While we also tested longer sequence, no significant performance improvement is observed.

In Fig. 10, both the mean value and the standard deviation of 10 trials by each algorithm are plotted. PSNR of the estimated parameter maps by all the algorithms increase as the sequence length increases. Notice here we do not include PSNR of the oracle estimator in Fig. 10 (d) for the visualization purpose (it ranges from 90 to 91 dB, far greater than the other algorithms). It can be seen that CSMRF+ML is stable and outperforms the other algorithms except for the density maps. MRF behaves not so well yet stably. It cannot effectively make use of the spatial information of the image and the $\mathcal{L}_2$ distance often fails to match query fingerprints to correct dictionary atoms. BLIP is better but unstable (i.e., having large variance) when the sequence length is short. This can be explained by the fact that its sampling strategy is independent each time and does not force to have different spatial encodings. This problem can be alleviated when a longer sequence is used. Furthermore, the PSNR of $T_1$, $T_2$ and $B_0$ maps by CSMRF+ML are close to the oracle estimator when the sequence length is no fewer than 500, which means that they are visually very close to the ground truth without obvious aliasing artifacts or noise.

### 3.4.3. Evaluation on sampling strategies

In order to test whether our proposed sampling scheme influences the performance of estimating multiple MR parameter maps, we compared the proposed sampling strategy with a baseline sampling strategy with an equivalent undersampling ratio of 4.69%. The baseline sampling strategy is to sample the k-space independently at each time, which follows a variable density random sampling pattern as the first sampling mask in our approach. It does not force choosing different rows of the k-space data at two consecutive time frames. All the experiments were repeated on 10 different slices. Each entry in Table 3 was obtained by averaging the results of the 10 slices.

In Table 3, we show the quantitative results of the proposed sampling strategy and those of the baseline strategy. The performance of the proposed sampling strategy was tested on both the MRF method and our proposed CSMRF+ML. We show that the baseline sampling strategy leads to not the best, yet stable results, which denotes that totally independent random variable sampling can guarantee a satisfactory performance.

The proposed sampling strategy results in better accuracy of parameter map estimation because our strategy increases the incoherence between the noise and the fingerprints. When c in Eq. (19) is 4, i.e., the probabilities of sampling the center 4 rows on the mask are fixed while other rows are dependent on the previous mask, the proposed sampling strategy achieved the best performance. Since our sampling ratio is 4.69%, only 12 out of 256 rows of the k-space data would be sampled. That means when $c$ is 8, there are only about 4 possible rows to choose from the high frequency parts of the k-space. On the contrary, when $c$ is 2, the low frequency parts are sampled too few. In both cases, the performance drops because of the inappropriate ratio of low frequency against high frequency.

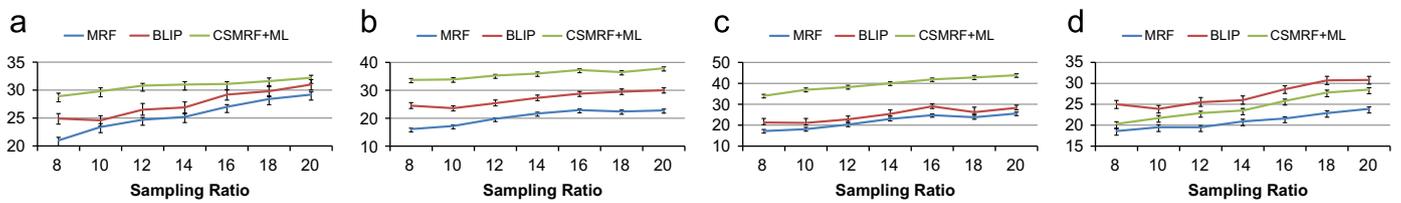

**Fig. 9.** Estimation accuracy vs. the number of rows sampled per slice. Both mean value and standard deviation are plotted. (a) PSNR for estimated $T_1$ maps with varying sampling ratios. (b) PSNR for estimated $T_2$ maps with varying sampling ratios. (c) PSNR for estimated off resonance frequency maps with varying sampling ratios. (d) PSNR for estimated proton density maps with varying sampling ratios.

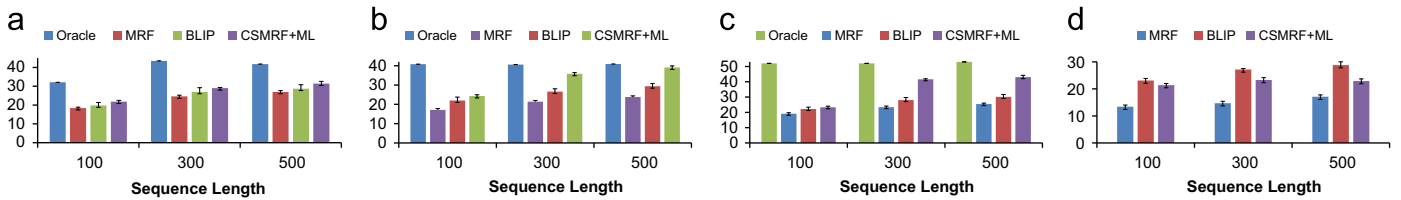

**Fig. 10.** PSNR vs. sequence length of all the tested algorithms. Both mean value and standard deviation are plotted. (a) PSNR of $T_1$ map vs. sequence length. (b) PSNR of $T_2$ map vs. sequence length. (c) PSNR of $B_0$ map vs. sequence length. (d) PSNR of proton density map vs. sequence length.

**Table 3**
Comparisons of the proposed sampling strategy with the baseline strategy. The baseline strategy undersamples the k-space independently at each time. Different $c$ values of our proposed sampling strategy were chosen to study how the sampling ratio between low frequency components and high frequency components would affect the results. The winning entries are marked in bold.

| | MRF | | | | | | | | CSMRF+ML | | | | | | | |
|---|---|---|---|---|---|---|---|---|---|---|---|---|---|---|---|---|
| Parameter type | $T_1$ | | $T_2$ | | $B_0$ | | proton density | | $T_1$ | | $T_2$ | | $B_0$ | | proton density | |
| Evaluation metric | PSNR | SSIM | PSNR | SSIM | PSNR | SSIM | PSNR | SSIM | PSNR | SSIM | PSNR | SSIM | PSNR | SSIM | PSNR | SSIM |
| Baseline | 24 | 0.921 | 20.6 | 0.831 | 22.9 | 0.864 | 20.6 | 0.822 | **33.9** | **0.981** | 31.2 | 0.983 | 39.0 | 0.979 | 22.1 | 0.808 |
| Proposed, $c=2$ | 26.4 | **0.952** | 21.8 | 0.852 | 24.3 | 0.892 | 21.8 | 0.864 | 29.7 | 0.967 | **35.9** | 0.98 | 32.4 | 0.955 | **27.7** | **0.873** |
| Proposed, $c=4$ | 26.5 | 0.943 | **21.3** | **0.852** | 22.5 | 0.881 | **21.9** | **0.862** | 30.0 | **0.975** | 35.6 | 0.977 | 39.9 | 0.985 | 24.7 | 0.872 |
| Proposed, $c=6$ | **27.9** | **0.947** | 20.1 | 0.825 | 20.3 | 0.845 | 21.2 | 0.864 | 30.1 | 0.969 | 35.2 | **0.982** | **41.8** | **0.991** | 25.2 | 0.866 |
| Proposed, $c=8$ | 26.6 | 0.945 | 20.6 | 0.839 | **23.7** | **0.899** | 20.4 | 0.855 | 30.7 | **0.975** | 34.6 | 0.982 | 38.1 | 0.983 | 23.4 | 0.851 |

### 3.4.4. Evaluation on distance metric learning methods

Our proposed algorithm can be combined with various distance metric learning algorithms. Here we focus on their performance on estimation of $T_1$ map (note that similar performance was observed for the $T_2$ and proton density maps). We compared the $\mathcal{L}_2$ distance with the distance metric learned by RCA [28], the Discriminative Component Analysis (DCA) [32] and the Local Fisher Discriminant Analysis (LFDA) [33]. The full-sized matrix is of $1000 \times 1000$ because the real part and the imaginary part of the training samples are concatenated. However, all these metric learning algorithms are able to learn a dimension-reduction transform $W \in \mathcal{R}^{T^- \times 1000}$, where $T^-$ is a number smaller than 1000. While we also tried different $T^-$ values for different algorithms, no significant difference was observed until $T^-$ is below 500, where the performance began to degrade. For a fair comparison on these metric learning algorithms, we learned a full-sized matrix for each of them.

In Table 4, we show the PSNR and SSIM of the parameter maps estimated by our algorithm with different metric learning algorithms. The performance of CSMRF with the $\mathcal{L}_2$ distance is listed as the baseline. DCA only improved the baseline a little while RCA performed best in our framework. The reason RCA works better than other distance metric learning algorithms might be that it does not force the samples with different labels to be far away from each other. This is consistent with our observations: a different sample may come from a neighboring dictionary atom, which is also a good approximation of the ground truth.

## 4. Discussions and conclusions

(1) *The learned distance metric*: The success of applying metric learning to MRF indicates that some dimensions may be more useful than others for matching MR Fingerprints to the dictionary atoms and there exist correlations between each dimension. Although learning a distance metric offline can discover important information in MR Fingerprints and may well tackle this problem, the collection of the ground truth data from phantoms or volunteers will take additional efforts. Moreover, calculating the Mahalanobis distances for each query fingerprint with all the dictionary atoms is more time consuming than their inner-products. Therefore, a better solution may be to specifically design the pulse sequence so that the MR Fingerprints of interest can be best distinguished with the inner-product.

(2) *Compressed sensing algorithm*: Currently the Conjugate Gradient descent with backtracking line search is used to optimize the proposed objective function. We will investigate more recent optimization methods for compressed sensing algorithms (e.g., [21,24]). Moreover, the compressed sensing step requires many empirically set parameters, such as the line search iterations and the step size. A possible research direction may be to design a systematic way of determining all such parameters.

**Table 4**
PSNR and SSIM of the estimated $T_1$ map with different distance metric learning algorithms.

| Distance metric | PSNR | SSIM |
| --- | --- | --- |
| $\mathcal{L}_2$ distance in [11] | 23.3 | 0.80 |
| RCA [28] | 32.3 | 0.98 |
| DCA [32] | 24.6 | 0.89 |
| LFDA [33] | 31.5 | 0.98 |

(3) *The proposed sampling strategy*: In our proposed sampling strategy, we empirically choose the $c$ value, which is equivalent to the number of rows near the k-space origin. We observe that when the sampling ratio is small (e.g. 8 out of 256 rows), the $c$ value should be closer to the total number of rows to be sampled. This is because in this case, if more low-frequency data (a larger $c$ value) are sampled, the image would be smoothed out and thus contains less noise. On the contrary, if more rows are allowed to be sampled (e.g., more than 32 out of 256 rows), then the $c$ value does not have to be too large. Generally, the users can set it according to the sampling ratio or by cross validation.

The proposed Cartesian-based sampling scheme in this study is quite different from the non-Cartesian sampling proposed for the original MRF study, and its implementation in various MR pulse sequences should be carefully considered in practice. For the normal spin-echo and gradient-echo pulse sequences widely used for clinical morphological imaging, the implementation of the proposed sampling scheme is highly practical because these sequences usually utilize phase-encoding gradient lobe and phase-encoding rewinding lobe pair prior and posterior to each echo acquisition. The frequency-encoding (or readout) gradient would not be lengthened. This implementation is also applicable for fast spin-echo sequences with multiple k-space row acquisition in each shot. The index sets for all time frames could be calculated once prior to acquisition and then applied to each time frame. Alternatively, the index set for each time frame $t-1$ could also be recorded for the calculation of the sampling mask for the next time frame $t$. It is worth noting that this Cartesian-based sampling scheme is technically challenging, so may not be suitable for echo planar imaging (EPI) sequence and gradient and spin echo (GRASE) sequence (either single-shot or multi-shot), in which gradient echo trains are used for frequency-encoding, and phase pre-winder lobe and small blip gradients are used for phase encoding. In these sequences, to achieve the proposed sampling mask, blips with different areas have to be used to skip some k-space rows due to the non-continuous k-space sampling. In this case, the large blip gradient lobes could inevitably prolong the required readout slope and hence the total gradient echo train duration, leading to more severe image distortion, SNR reduction and many other artifacts such as ghosting. Nevertheless, the implementation of our proposed method involves tremendous efforts in pulse sequence development and its performance on prospectively undersampled real MRI data has to be thoroughly validated in future works.

The main difference between CSMRF-ML and BLIP is that we learned the distance metric from the data instead of a pre-defined one. This new metric allows us to better match the fingerprints to the dictionary atoms. Besides, we explicitly ask the reconstructed images to be sparse in some transform domain while BLIP tried to apply the sparse prior on the proton density maps and found no significant improvement. We also proposed a variable density randomized sampling strategy while BLIP adopted a uniform sampling strategy.

The MRF method is a new approach to magnetic resonance and not fully exploited yet. In this work, we propose a compressed sensing framework for MRF with distance metric learning. A novel algorithm is proposed to reconstruct the undersampled data and estimate the MR parameters. It first solves the compressed sensing optimization problem and then projects the signal evolution to the Bloch response manifold with a learned distance metric. Thus the solution benefits from both temporal and spatial regularization. A novel sampling strategy is also proposed for maximizing the incoherence between the fingerprint and the aliasing error on it. We conducted numerical simulations to demonstrate the effectiveness of our framework. When compared with MRF [11] and BLIP [26], our algorithm outperforms them in terms of accuracy of parameter map estimation.

## Acknowledgment

This work was supported in part by Lui Che Woo Institute of Innovative Medicine (No. LCWIM 8303122), in part by NSFC (No. 61301269), in part by the PhD Programs Foundation of MOE of China (No. 20130185120039), in part by China Postdoctoral Science Foundation (No. 2014M552339) and in part by Sichuan High Tech R&D Program (No. 2014GZX0009).

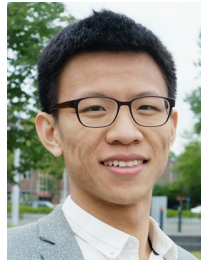

**Qinwei Zhang** is a PhD candidate at University of Amsterdam in The Netherlands. He is working as a research assistant in the Academic Medical Center, Amsterdam. His research is focused on Cardiovascular MRI, diffusion imaging on 3T and 7TMRI.

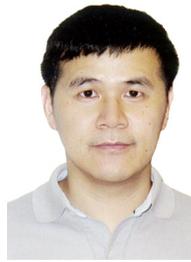

**Jing Yuan** obtained his bachelor's and master's degrees from Tsinghua University, and PhD degree from the University of Hong Kong, all majored in Electronic Engineering. He was a postdoctoral research fellow in Harvard Medical School and an assistant professor in the Department of Imaging and Interventional Radiology, Chinese University of Hong Kong. Now he is a research fellow and MRI physicist of medical physics and research department, Hong Kong Sanatorium & Hospital. Dr. Yuan has intensive experiences on technical and clinical MRI research for over 10 years and his current research interest is MRI-aided radiation therapy.

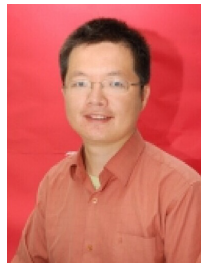

**Xiaogang Wang** received the BS degree from the University of Science and Technology of China in 2001, the MS degree from the Chinese University of Hong Kong in 2003, and the PhD degree from the Computer Science and Artificial Intelligence Laboratory at the Massachusetts Institute of Technology in 2009. He is currently an Associate Professor in the Department of Electronic Engineering at The Chinese University of Hong Kong. His research interests include computer vision.

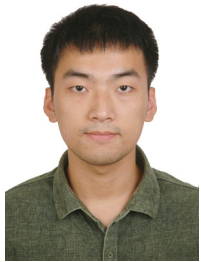

**Zhe Wang** received the BEng from the Department of Optical Engineering of Zhejiang University, China, in 2012. He is currently working toward a PhD degree in the Department of Electronic Engineering, the Chinese University of Hong Kong. His research interest is focused on compressed sensing, magnetic resonance imaging, image segmentation and object detection.

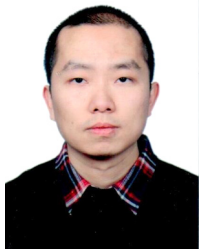

**Hongsheng Li** received the bachelor's degree in automation from East China University of Science and Technology, and the master's and doctorate degrees in computer science from Lehigh University, Pennsylvania, in 2006, 2010, and 2012, respectively. He is a research assistant professor in the Department of Electronic Engineering at the Chinese University of Hong Kong. His research interests include computer vision, medical image analysis and machine learning.